\documentclass[final,1p,times]{elsarticle}

 \usepackage{graphics}

\usepackage{amssymb}

\begin{document}

\begin{frontmatter}



\title{Fermi acceleration and suppression of Fermi acceleration in a time-dependent Lorentz Gas.}


\author{Diego F. M. Oliveira}

\address{Max Planck Institute for Dynamics and Self-Organization -- Bunsenstrasse 10 - D-37073 - G\"ottingen - Germany \\ CAMTP - Center For Applied Mathematics and Theoretical Physics -- University of Maribor - Krekova 2 - SI-2000 - Maribor - Slovenia.}

\ead{diegofregolente@gmail.com}

\author{ J\"urgen Vollmer}

\address{Max Planck Institute for Dynamics and Self-Organization -- Bunsenstrasse 10 - D-37073 - G\"ottingen - Germany.}

\ead{juergen.vollmer@ds.mpg.de}

\author{ Edson D. Leonel}

\address{Departamento de Estat\'{\i}stica, Matem\'atica Aplicada e
Computa\c c\~ao -- Instituto de Geoci\^encias e Ci\^encias Exatas --
Universidade Estadual Paulista -- \\ Av. 24A, 1515 -- Bela Vista -- CEP:
13506-900 -- Rio Claro -- SP -- Brazil.}

\ead{edleonel@rc.unesp.br}

\begin{abstract}
We study some dynamical properties of a Lorentz gas. We have
considered both the static and time dependent boundary. For the
static case we have shown that the system has a chaotic component
characterized with a positive Lyapunov Exponent. For the
time-dependent perturbation we describe the model
using a four-dimensional nonlinear map. The behaviour of the average velocity is considered in two situations
(i) non-dissipative and (ii) dissipative. Our results show that the
unlimited energy growth is observed for the non-dissipative case. However, when dissipation, via damping coefficients, is introduced
the senary changes and the unlimited engergy 
growth is suppressed. The behaviour of the average velocity is described using scaling approach.

\end{abstract}

\begin{keyword}
Billiard, Lorentz Gas, Lyapunov exponents, Fermi Acceleration, Scaling.

\end{keyword}

\end{frontmatter}


\section{Introduction}
\label{sec1} 
The process in which a classical particle acquies unlimited
energy growth from collisions with heavy boundaries, also called as phenomenon of Fermi acceleration,
 were first reported by Enrico Fermi \cite{ref1} as an
attempt to explain the acceleration of cosmic rays. He proposed that
such behaviour was due to interaction between charged particles and
time-dependent magnetic fields. Soon after \cite{ref1} some
alternative models have been proposed using different approaches with application in different fields of science
including molecular physics \cite{refmole}, optics \cite{refopt}, nanostructures \cite{refnano}, quantum dots \cite{refdots}.

One of the most studied versions of the problem is the well known
one-dimensional Fermi-Ulam model (FUM)
\cite{ref2,ref3,ref4,ref5,ref6,ref7,ref8}. The model consists of a
classical particle confined and bouncing between two rigid walls in
which one of them is fixed and the other one moves according to a
periodic function. It is well known that the phase space, in the absence of dissipation, shows a
mixed structure in the sense that depending on the combinations of
control parameters and initial conditions, both invariant spanning curves (also called as invariant tori),
 chaotic seas and Kolmogorov- Arnold-Moser (KAM) islands are
all observed. An alternative model was latter proposed by
Pustylnikov \cite{refpust1,refpust2}. Such system consists of a
classical particle bouncing in a vertical moving platform under the
effect of an external constant gravitational field
\cite{ref9,ref10,ref11,ref12,ref13,ref14,ref15}. Both models seems
to be quite similar. However there are many differences between
them. The main difference is, that in the  FUM framework the Fermi
acceleration is not observed. On the other hand, for specific
combinations of both control parameters and initial conditions the
phenomenon of unlimited energy growth can be observed in the bouncer
model. This apparent contradictory result was latter discussed and
explained by Lichtenberg and Lieberman \cite{reflili1,reflili2} and
can be easily understood by looking at the space phase. The FUM has
a set of invariant spanning curves limiting the size of the chaotic
sea (as well as the particle's velocity), but such invariant tori,
which could be interpreted as a barrier,
 they are not observed in the bouncer model and the energy grows
unbounded. However, in two-dimensional systems it is not so simple to say if the phenomenon of Fermi acceleration will be observed or not. In
this sense  a conjecture was proposed by Loskutov-Ryabov-Akinshin (LRA) \cite{ref16}. Such conjecture, known as LRA-conjecture, 
states that a chaotic componet
in the phase space with static boundary is a sufficient condition to observe Fermi acceleration when  a perturbation is introduced. Results 
that corroborate the validity of this conjecture include the time dependent oval billiard \cite{refov}, stadium billiard \cite{refsta}.

When dissipation is introduce, it causes a drastic change on the phase space. Invariant spanning curves are destroyed.
the elliptic fixed points turns into sinks and the chaotic sea is replaced by a chaotic attractor \cite{refcrisis}. However, the influence 
of dissipation on the averave velocity is still not fully understood. It was discussed by Leonel \cite{refbrea} some results 
for a one-dimensional
Fermi-Ulam model. It is well known that such system in its original
formulation the particle do not have unlimited energy growth. If
one consider the motion of the moving wall to be random the phenomenon of Fermi acceleration is observed. However, the introduction of
inelastic collision is enough to suppress Fermi acceleration. Results considering also the one dimension Fermi-Ulam model
under an external force of type sawtooth \cite{refladeira} under effects of dissipation generated from a sliding of a body against a 
rough surface has also been considered. The external perturbation of sawtooth type was chosen because the oscillating wall always gives 
energy to the particle after collisions. The main question is: Would it be possibel to suppress Fermi acceleration under the efect of 
 dissipation generated from a sliding of a body against a rough surface?
The answer is not so simple and it dependes on both, initial conditionand the combination of control parameters. It was observed that 
for a certain range of parameters Fermi acceleration indeed happen, however, it is suppresses under specific conditions.

In this paper, we study the problem of a Lorentz Gas considering
both the static and the time-dependent boundary. In the first part
of our work we study the Lorentz Gas with static boundary. We
derived a two dimensional nonlinear mapping that describes the
dynamics of the model. We obtain the phase space and we show that it
is chaotic with positive Lyapunov Exponent. Then, in the second part
of our paper we introduce a time-dependent perturbation on the
boundary. There are many different ways to introduce time dependent
perturbation and the most common methods are: stochastic case, where
the boundary changes according to a random function \cite{ref17} ;
regular case, where the size of the boundary varies according to an
harmonic law \cite{ref18}. However, in both situations the center of
mass it is assumed to be fixed. Now, for the first time, we
introduce a different kind of time dependent perturbation for a
Lorentz gas. We assume that the radius of the scatters are fixed and
the center of mass changes according to an harmonic function. Our
main goal in this part of our work is to verify the validity of the
LRA conjecture, which is confirmed when we studied the behaviour of
the average velocity for an ensemble of particles. Since the phenomenon of Fermi Acceleration is present in this model our next approach
is to introduce dissipation into the model via damping coefficients and trying to understand what is the influence of dissipation on 
the particle’s behaviour. Our results allow us to confirm that when inelastic collisions is introduced into the model
 it is a sufficient condition to break down the phenomenon of Fermi
acceleration. In both, conservative as well as dissipative case, we describes the behaviour of average velocity using scaling formalism.

The paper is organized as follows. In section \ref{sec2} we describe
how to obtain the two-dimensional mapping that describes the
dynamics of the static system. Section \ref{sec3} is devoted to
discuss the time dependent model as well as our numerical results.
Finally, conclusion and acknowledgments are drawn in section
\ref{sec4}.

\section{A static Lorentz Gas and the mapping}
\label{sec2}

\begin{figure}[t]
\centerline{\includegraphics[width=0.60\linewidth]{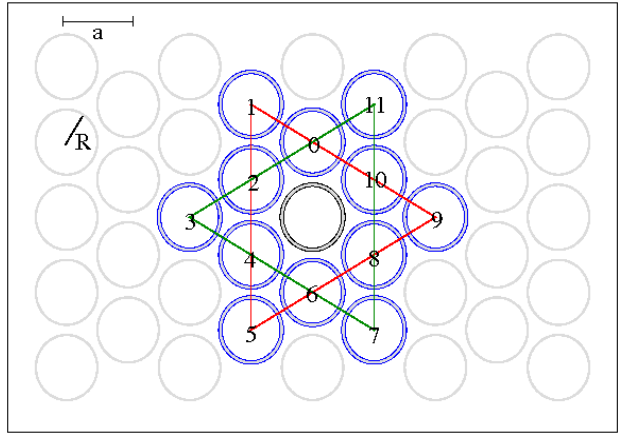}}
\caption{\it{Illustration of the Lorentz gas with triangular configuration.}}
\label{fig1}
\end{figure}

\begin{figure}[t]
\centerline{\includegraphics[width=0.60\linewidth]{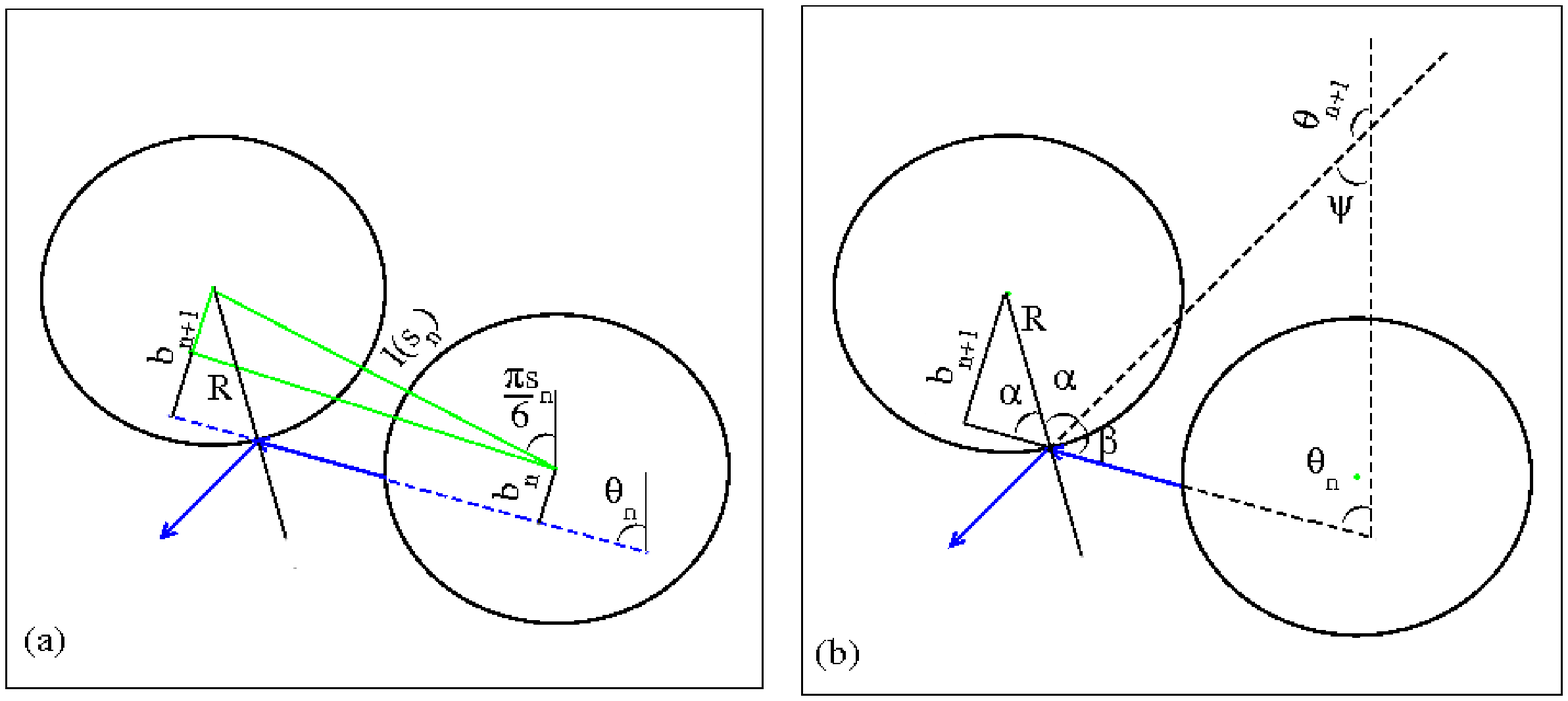}}
\caption{\it Dependence of {(a) $b_{n+1}$ on $b_n$ and $\theta_n$; (b) $\theta_{n+1}$ on $b_{n+1}$ and $\theta_n$.}}
\label{fig2}
\end{figure}

In this section we discuss all the details needed for the
construction of a non-linear mapping that describes the dynamics of
the problem. The model consists of a classical particle of mass $m$
suffering elastic collisions with circular scatters (see Fig
\ref{fig1}). We choose a triangular arrangement of the scatters
(which also seems to be the Star of David if one connect circles 1,5
and 9 and 11,7 and 3) in order to avoid particles traveling
infinitely far between collision and the fixed lattice spacing {\it
a} to be twice the radius of the scatters, {\it R=a/2}. The system
is described in terms of a two dimensional mapping
$\Xi(\theta_n,b_n)=(\theta_{n+1},b_{n+1})$ where the dynamical
variable $\theta_n$ denotes the direction of the trajectory while
$b_n$ is the impact parameter. Given an initial condition
$(\theta_0,b_0)$, the particle starts from the black circle (center
in Fig. \ref{fig1}) and hit one of the 12 others circles. In this
sense, we specify the scattered hit in the collision {\it n+1} by
$s_n=0,...,11$ and we introduce $l(s_n)$ for the distance of this
scatterer and the scatterer hit at the collision {\it n+1}. $l(s_n)$
can assume the values of $2a/\sqrt{3}$ and $2a$ for even and odd
values of $s_n$, respectively. Additionally, when the particle hits
the boundary it is specularly reflected with the same absolute
velocity. The particle does not suffer influences of any external
field along its linear trajectory.  From the green triangle in Fig
\ref{fig2} (a) one can  easily verify

\begin{eqnarray}
{\sin(\theta_n-{{\pi {s_n}\over 6}})}=-{(b_{n+1}-b_n)\over l(s_n)}.
\label{eq1}
\end{eqnarray}

Moreover, from Fig. \ref{fig2} (b) one can find that

\begin{eqnarray}
\alpha&=& \arcsin({-b_{n+1}/ R}), \\ \beta&=&\pi-2\alpha, \\ \psi&=&2 \alpha -\theta_n, \\ \theta_{n+1}&=&\pi - 2\alpha+\theta_n.
\label{eq2}
\end{eqnarray}

Such a result allow us to obtain $\theta_{n+1}$,

\begin{eqnarray}
\theta_{n+1}=\pi +\theta_n +2\arcsin({b_{n+1}/ R}).
\label{eq3}
\end{eqnarray}

From Eq. \ref{eq1} it is easily to find that the impact parameter,
$b_{n+1}$, is given by

\begin{eqnarray}
b_{n+1}=b_n -l(s_n){\sin(\theta_n-{{\pi {s_n}\over 6}})}.
\label{eq4}
\end{eqnarray}

Thus, the mapping that describe the dynamics of a two dimensional Lorentz gas is given by

\begin{equation}
\Xi :\left\{\begin{array}{ll}
\theta_{n+1}=\pi +\theta_n +2\arcsin({b_{n+1} \over R})~~\\
b_{n+1}=b_n -l(s_n){\sin(\theta_n-{{\pi {s_n}\over 6}})}\\
\end{array}
\right.~,
\label{eq5}
\end{equation}
by definition $b\in [-R,R]$ and $\theta\in [0, 2\pi]$  is a
counterclockwise angle such that $\Xi$ is defined on the fundamental
domain $[-R,R]\times[0,2\pi]$. From the mapping $\Xi$, Eq.
(\ref{eq5}), one can easily obtain the Jacobian Matrix, $J$, which
is defined as

\begin{equation}
J=\left(\begin{array}{ll}
{{\partial \theta_{n+1}}\over{\partial \theta_n}}~&~{\partial
\theta_{n+1}\over{\partial b_n}}\\
{{\partial b_{n+1}}\over{\partial \theta_n}}~&~{\partial
b_{n+1}\over{\partial b_n}}\\
\end{array}
\right),
\label{Eq211}
\end{equation}
with coefficients given by
\begin{eqnarray}
{\partial \theta_{n+1} \over \partial \theta_n}&=&1-2\sqrt{R^2-b^2_{n+1}}l(s_n)\cos(\theta_n-{{\pi {s_n}\over 6}}) ~ , \\
{\partial \theta_{n+1} \over \partial b_n}&=&2\sqrt{R^2-b^2_{n+1}}~ ,\\
{\partial b_{n+1} \over \partial \theta_n}&=&-l(s_n)\cos(\theta_n-{{\pi {s_n}\over 6}})~,\\
{\partial b_{n+1} \over \partial b_n}&=&1~.
\label{Eq212}
\end{eqnarray}

After some easy calculation one can show that the mapping $\Xi$ preserves the phase space measure since $\det(J)=1$.

\begin{figure}[t]
\centerline{\includegraphics[width=0.60\linewidth]{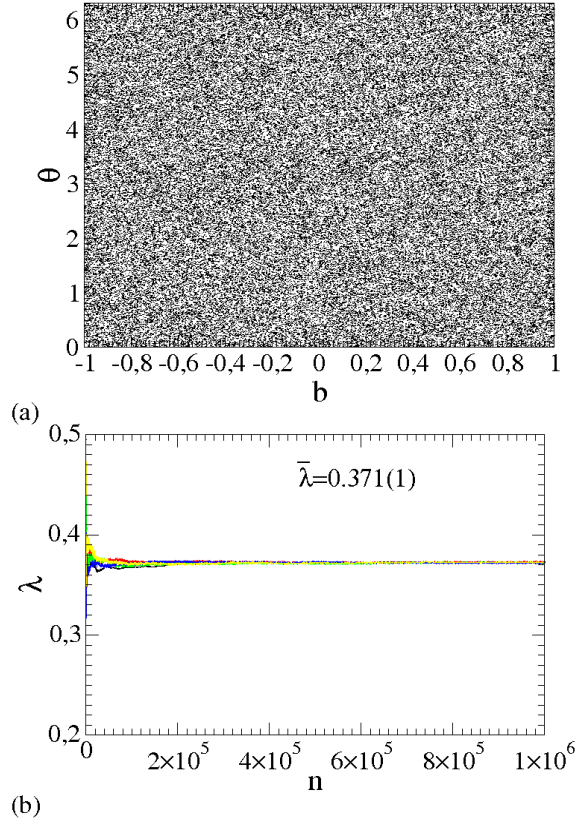}}
\caption{(a) Phase space generated from iteration of mapping (\ref{eq5}); (b) Behaviour of the positive Lyapunov
exponent of the chaotic sea. The control parameter used in both figures were $a=2$.}
\label{figlyap}
\end{figure}

It is well know the Lyapunov exponents are  an important tool to
identify whether the model is chaotic or not. As discussed in
\cite{ref20}, the Lyapunov exponents are defined as
\begin{equation}
\lambda_j=\lim_{n\rightarrow\infty}{1\over{n}}\ln|\Lambda_j|~~,~~j=1,
2~~,
\label{eql4}
\end{equation}
where $\Lambda_j$ are the eigenvalues of
$M=\prod_{i=1}^nJ_i(\theta_i,b_i)$ and $J_i$ is the Jacobian matrix
evaluated over the orbit $(\theta_i,b_i)$. However, a direct
implementation of a computational algorithm to evaluate Eq.
(\ref{eql4}) has a severe limitation to obtain $M$. Even in the
limit of short $n$, the components of $M$ can assume very different
orders of magnitude for chaotic orbits and periodic attractors
yielding impracticable the implementation of the algorithm. In order
to avoid such problem we note that $J$ can be written as $J=\Theta
T$ where $\Theta$ is an orthogonal matrix and $T$ is a right
triangular matrix. Thus we rewrite $M$ as $M=J_nJ_{n-1}\ldots
J_2\Theta_1\Theta_1^{-1}J_1$, where $T_1=\Theta_1^{-1}J_1$. A
product of $J_2\Theta_1$ defines a new $J_2^{\prime}$. In a next
step, it is easy to show that $M=J_nJ_{n-1}\ldots
J_3\Theta_2\Theta_2^{-1}J_2^{\prime}T_1$. The same procedure can be
used to obtain $T_2=\Theta_2^{-1}J_2^{\prime}$ and so on. Using this
procedure the problem is reduced to evaluate the diagonal elements
of $T_i:T_{11}^i,T_{22}^i$. Finally, the Lyapunov exponents are now
given by
\begin{equation}
\lambda_j=\lim_{n\rightarrow\infty}{1\over{n}}\sum_{i=1}^n
\ln|T_{jj}^i|~~ , ~~j=1,2~~.
\label{eql5}
\end{equation}
If at least one of the $\lambda_j$ is positive then the orbit is
classified as chaotic. Additionally, in conservative systems
$\lambda_1+\lambda_2=0$ and in dissipative systems
$\lambda_1+\lambda_2<0$. The phase space for the mapping (\ref{eq5})
is shown in Fig. \ref{figlyap}(a). For the same control parameters
used in Fig. \ref{figlyap} (a), $a=2$, we have also evaluated
numerically the positive Lyapunov exponent as one can see in Fig.
\ref{figlyap}(b). The average of the positive Lyapunov exponent for
the ensemble of the 5 time series gives $\bar{\lambda}=0.371\pm
0.001$ where the value $0.001$ corresponds to the standard deviation
of the five samples.

\section{A Time Dependent Lorentz Gas.}
\label{sec3}

In this section we introduce a new kind of time dependent perturbation in a Lorentz gas. We assume that the radius of the scatters are
fixed and the center of
mass changes according to an harmonic function $f(t)$, in particular we assume the case where

\begin{eqnarray}
f_i(t)=\epsilon_i[1+\cos(t)], ~~~~ for~~ i=x,y.
\label{eq34}
\end{eqnarray}
where $\epsilon_i$ is the amplitude of the time-dependent perturbation and $t$ is the time. When we introduce time dependent perturbation into
the model, two new dynamical variables appears,
namely, velocity and time. Now, we have a two dimensional system described in terms of a four dimension nonlinear mapping which relates the
collision $n^{th}$ with the $(n+1)^{th}$, i.e.,
 $\Xi(\theta_n,b_n,V_n,t_n)=(\theta_{n+1},b_{n+1},V_{n+1},t_{n+1})$. The corresponding
variables are: the direction of the trajectory, $\theta_n$; the impact parameter, $b_n$; the absolute velocity of the particle, $V_n$ and
the instant of the hit with the boundary,
$t_n$.

Assuming that an initial condition $(\theta_0,b_0,V_0,t_0)$ is
given, we can obtain the equation that describe the dynamics of the
system. Thus, according to our construction,
 the cartesian components of $R$
are given by
\begin{eqnarray}
X(\delta_n,t_n)&=&R \cos(\delta_n)+\epsilon_x [1+\cos(t_n)], \\
Y(\delta_n,t_n)&=&R \sin(\delta_n)+\epsilon_y [1+\cos(t_n)].
\label{eq35}
\end{eqnarray}
where $\delta_n$ is the angular position which is given by $\delta_n=\pi/2+\theta_n-\arcsin({b_n}/R)$. Since we already know the angle that
the particle's trajectory does with the horizontal $(\theta_n+\pi/2)$ and the position of the hit at
the collision $n^{th}$, we can obtain the vector velocity of the particle that is written as
\begin{eqnarray}
\overrightarrow{V}_n=\vert\overrightarrow{V_n}\vert
[\cos(\theta_n+\pi/2)\widehat{i}+\sin(\theta_n+\pi/2)\widehat{j}]~,
\label{eq36}
\end{eqnarray}
where $\widehat{i}$ and $\widehat{j}$ represent the unit vectors with
respect to the X and Y axis, respectively. The above expressions allow us to obtain the
position of the particle as a function of time for $t \geq t_n$:
\begin{eqnarray}
X_{p}(t)&=&X(\delta_n,t_n)+\vert\overrightarrow{V}_n\vert\cos(\theta_n+\pi/2)(t-t_n)~,\\
Y_{p}(t)&=&Y(\delta_n,t_n)+\vert\overrightarrow{V}_n\vert\sin(\theta_n+\pi/2)(t-t_n)~.
\label{eq37}
\end{eqnarray}

The index $p$ denotes the corresponding coordinates of the particle.
In order to know the position of the particle at $(n+1)^{th}$
collision we need to solve numerically the following equation

\begin{eqnarray}
r=\sqrt{[X_x(t)-X_{p}(t)]^2+[Y_y(t)-Y_{p}(t)]^2}\cong R~,
\label{eq38}
\end{eqnarray}
where both $X_x$ and $Y_y$ are given by
\begin{eqnarray}
X_x(t)=l_x+\epsilon_x[1+\cos(t)], \\
Y_y(t)=l_y+\epsilon_y[1+\cos(t)]~,
\label{eq39}
\end{eqnarray}
being $l_x$ and $l_y$ the X and Y components of $l(s_n)$; this
distance is measured from the origin of the coordinates system to
the center of the $s_n=0,...,11$ scatters at $(n+1)^{th}$ collision.
Since we already know the position of the particle at the collision
$(n+1)^{th}$, one can easily find the distance between two
successive impacts, which is given by
$d=\sqrt{[X_{p}(t)-X(\delta_n,t_n)]^2+[Y_{p}(t)-Y(\delta_n,t_n)]^2}.$
Then, the time at $(n+1){th}$ collision is obtained evaluating the expression
\begin{eqnarray}
t_{n+1}=t_n+ {{\sqrt{[X_{p}(t)-X(\delta_n,t_n)]^2+[Y_{p}(t)-Y(\delta_n,t_n)]^2}} \over \vert\overrightarrow{V}_n\vert}~.
\label{eq41}
\end{eqnarray}

The next step is to obtain the impact parameter, $b_{n+1}$, which is given by

\begin{eqnarray}
b_{n+1}=b_n -l(s_n){\sin\left(\theta_n-\psi+{{\pi}\over 2}\right)},
\label{eq42}
\end{eqnarray}
where $l(s_n)=\sqrt{(\Delta X)^2+(\Delta Y)^2}$ and $\psi=\arctan(\Delta X / \Delta Y)$ with
\begin{eqnarray}
\Delta X=l_x+\epsilon_x[\cos(t_{n+1})-\cos(t_n)]\\
\Delta Y=l_y+\epsilon_y[\cos(t_{n+1})-\cos(t_n)]
\label{eq42_a}
\end{eqnarray}

Moreover, the new direction of the trajectory, $\theta_{n+1}$ is
\begin{eqnarray}
\theta_{n+1}=\pi +\theta_n +2\arcsin\left[{{b_{n+1} \over R}}\right].
\label{eq43}
\end{eqnarray}

We already know $(\theta_{n+1},b_{n+1},t_{n+1})$, however we still have to find $\overrightarrow{V}_{n+1}$. At the new angular
position $\delta_{n+1}$, the
unitary tangent and normal vectors are
\begin{eqnarray}
\overrightarrow{T}_{n+1}&=&\cos(\delta_{n+1})\widehat{i}+\sin(\delta_{n+1}
)\widehat{j}~,\\
\overrightarrow{N}_{n+1}&=&-\sin(\delta_{n+1})\widehat{i}+\cos(\delta_{n+1}
)\widehat{j}~.
\label{eq45}
\end{eqnarray}

Since the referential frame of the boundary is moving, then, at the instant of the collision, according to our construction, the following
conditions must be matched
\begin{eqnarray}
\overrightarrow{V^\prime}_{n+1}\cdot\overrightarrow{T}_{n+1}
&=&\gamma\overrightarrow{V^\prime}_{n}\cdot\overrightarrow{T}_{n+1}~,
\label{eq52}\\
\overrightarrow{V^\prime}_{n+1}\cdot\overrightarrow{N}_{n+1}
&=&-\delta\overrightarrow{V^\prime}_{n}\cdot\overrightarrow{N}_{n+1}~,
\label{eq53}
\end{eqnarray}
where $\gamma \in [0,1]$ and $\delta \in [0,1]$ are damping coefficients, which means that the particle can lose velocity/energy upon collision. 
The complete inelastic case occurs when
$\gamma=\delta=0$. On the other hand, when $\gamma=\delta=1$ corresponds to the conservative case. The upper prime indicates that the velocity of the particle is measured with respect to the moving boundary
referential frame. 

Hence, one can easily find that
\begin{eqnarray}
\overrightarrow{V}_{n+1}\cdot\overrightarrow{T}_{n+1}
&=&\gamma\overrightarrow {V}_{n}\cdot\overrightarrow{T}_{n+1}
+(1-\gamma)\overrightarrow{
V}_{b}(t_{n+1})\cdot\overrightarrow{T}_{n+1}~.
\label{eq54}
\end{eqnarray}
\begin{eqnarray}
\overrightarrow{V}_{n+1}\cdot\overrightarrow{N}_{n+1}
&=&-\delta\overrightarrow{V}_{n}\cdot\overrightarrow{N}_{n+1}
+(1+\delta)\overrightarrow{V}_{b}(t_{n+1}
)\cdot\overrightarrow {N}_{n+1}~,
\label{eq55}
\end{eqnarray}
where $\overrightarrow{V}_{b}(t_{n+1})$ is the velocity of the boundary which is written
as
\begin{eqnarray}
\overrightarrow{V}_{b}(t_{n+1})=-\sin(t_{n+1})[\epsilon_x\widehat{i}+\epsilon_y\widehat{j}]~,
\label{eq56}
\end{eqnarray}

Finally, the velocity at $(n+1)^{th}$ collision is given by
\begin{eqnarray}
|\overrightarrow{V}_{n+1}|=\sqrt{(\overrightarrow{V}_{n+1}\cdot\overrightarrow{T}_{n+1}
)^2+(\overrightarrow{V}_{n+1}\cdot\overrightarrow{N}_{n+1})^2}~.
\label{eq58}
\end{eqnarray}

\subsection{Numerical Results}
Our numerical results for the time-dependent Lorentz Gas shows
basically the behaviour of the average velocity of the particle. Two different procedures were
applied in order to obtain the average velocity. Firstly, we
evaluate the average velocity over the orbit for a single initial
condition which is defined as
\begin{eqnarray}
{V}_i={{1}\over{n+1}}\sum_{j=0}^nV_{i,j}~,
\label{eq015}
\end{eqnarray}
where the index $i$ corresponds to a sample of an ensemble of initial
conditions. Hence, the average velocity is written as
\begin{eqnarray}
\overline{V}={{1}\over{M}}\sum_{i=1}^MV_i~,
\label{eq16}
\end{eqnarray}
where $M$ denotes the number of different initial conditions. We
have considered $M=1000$ in our simulations and from now on, we also
fixed the value $a=2$.

\begin{figure}[t]
\centerline{\includegraphics[width=0.60\linewidth]{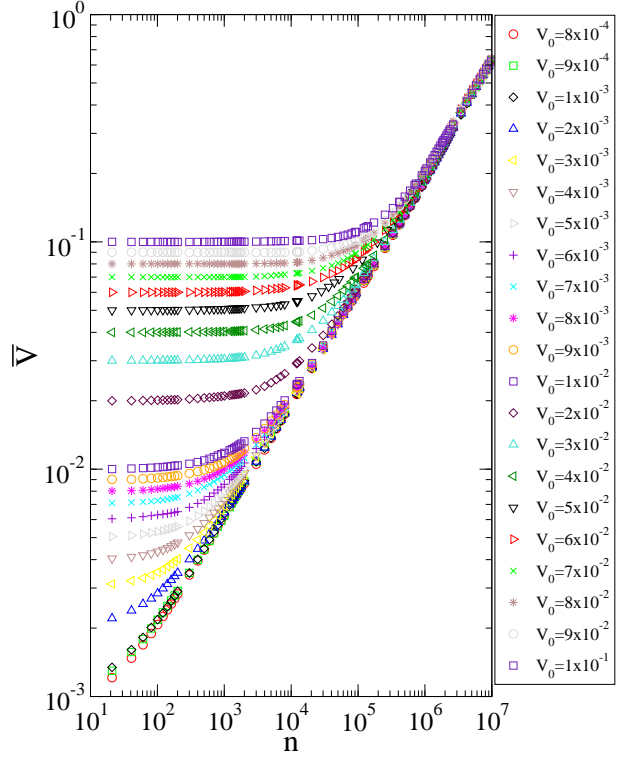}}
\caption{\it{Behaviour of ${\bar V}\times n$ for different initial
velocities. The control parameters used were $\epsilon_x=10^{-4}$, $\epsilon_y=3\times1 0^{-4}$ and $a=2$}}
\label{fig19}
\end{figure}

\subsection{Scaling results for the conservative case}

Our main goal in this section is describe a scaling present in the model
for the conservative case, where, in Eq. \ref{eq52} and Eq. \ref{eq53}, we assume $\gamma=\delta= 1$,
and also verify the validity of LRA conjecture.

 We begin discussing a scaling observed for the average velocity of the
particle as function of $V_0$ and $n$. It is shown in Fig.  \ref{fig19} the behavior of the $\bar V\times
n$ for different initial velocities. Hence, the control parameters
used in Fig. \ref{fig19} are $\epsilon_x=10^{-4}$,
$\epsilon_y=3\times1 0^{-4}$. Additionally, 21 different values of
$V_0$ were chosen and for each one we randomly chose $t \in [0,
2\pi]$, $\theta \in [0, 2\pi]$ and $b \in
[1-(\epsilon_x+\epsilon_y), -1+(\epsilon_x+\epsilon_y)]$. As one can
see, all curves of the $\bar V$ behave quite similarly in the sense
that: for short $n$, the average velocity remains constant, then
after a changeover, all the curves start growing with the same
exponent. Such behaviour is typical in systems that can be described
using scaling approach. Based on the behavior shown in Fig.
\ref{fig19}, we propose the following hypotheses:
\begin{enumerate}
\item{When $n\ll{n_x}$, $\bar V$ behaves according to
\begin{equation}
\bar{V}_{sat}\propto V_{0}^{\zeta}~,
\label{eq613}
\end{equation}
}
\item{For $n \gg n_x$, the average velocity is given by
\begin{equation}
\bar{V}\propto {n}^{\nu}~,
\label{eq614}
\end{equation}
}

\item{The crossover iteration number that marks the change from constant
velocity to growth is written as
\begin{equation}
n_x\propto V_0^{\xi}~,
\label{eq615}
\end{equation}
where $\zeta$ and $\nu$ are the critical exponents and $\xi$ is a
dynamic exponent. }
\end{enumerate}

After consider these three initial suppositions, we suppose that the average
velocity is described in terms of a generalized homogeneous function of the type
\begin{equation}
\bar{V}(V_0,n)={\rm l}\bar{V}({\rm l}^p{V_0},{\rm l}^q{n})~,
\label{eq616}
\end{equation}
where ${\rm l}$ is the scaling factor, $p$ and $q$ are scaling
exponents that in principle must be related to $\zeta$, $\nu$ and $\xi$. If
we chose ${\rm l}^{p}V_0=1$, then ${\rm l}=V_0^{-1/p}$ and Eq.
(\ref{eq616}) is given by
\begin{figure}[t]
\centerline{\includegraphics[width=0.60\linewidth]{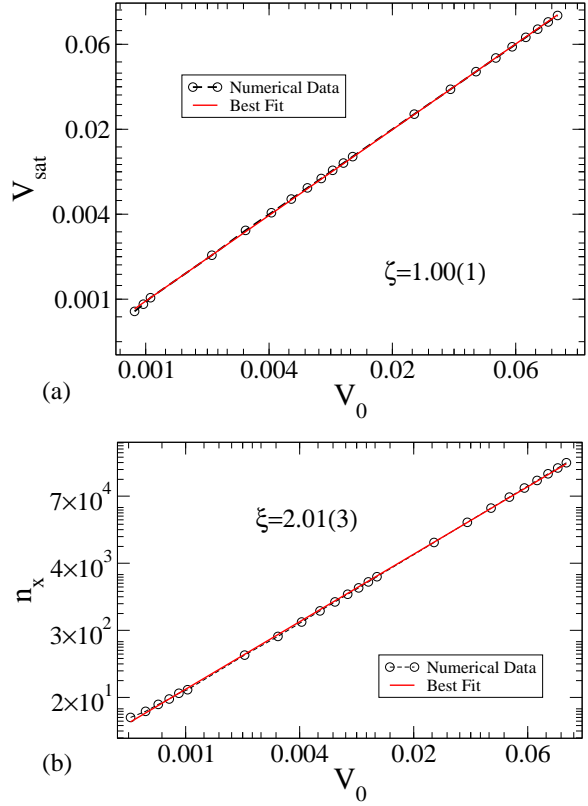}}
\caption{\it{(a) Plot of $V_{sat} \times V_0$. (b) Behaviour of $n_x$ as
function of $V_0$.}}
\label{fig20}
\end{figure}
\begin{equation}
\bar{V}(V_0,n)={V_0}^{-1/p}\bar{V}_1(V_0^{-q/p}n)~,
\label{eq617}
\end{equation}
where $\bar{V}_1(V_0^{-q/p}n)=\bar{V}(1,V_0^{-q/p}n)$ is assumed to be
constant for $n\ll{n_x}$. Comparing Eq. (\ref{eq617}) and Eq.
(\ref{eq613}), we obtain $\zeta=-1/p$.

On the other hand, Choosing now ${\rm l}=n^{-1/q}$, Eq. (\ref{eq616}) is
rewritten as
\begin{equation}
\bar{V}(V_0,n)=n^{-1/q}\bar{V}_2(n^{-p/q}V_0)~,
\label{eq618}
\end{equation}
where the function $\bar{V}_2$ is defined as
$\bar{V}_2(n^{-p/q}V_0)=\bar{V}(n^{-p/q}V_0,1)$. It is also assumed to
be constant for $n\gg{n_x}$. Comparing Eq. (\ref{eq618}) and Eq.
(\ref{eq614}) we find $\nu=-1/q$. Given the two different expressions of the
scaling factor {\rm l}, we obtain a relation for the dynamic exponent $\xi$, which is given by
\begin{equation}
\xi={\zeta \over \nu}~.
\label{eq619}
\end{equation}
\begin{figure}[t]
\centerline{\includegraphics[width=0.60\linewidth]{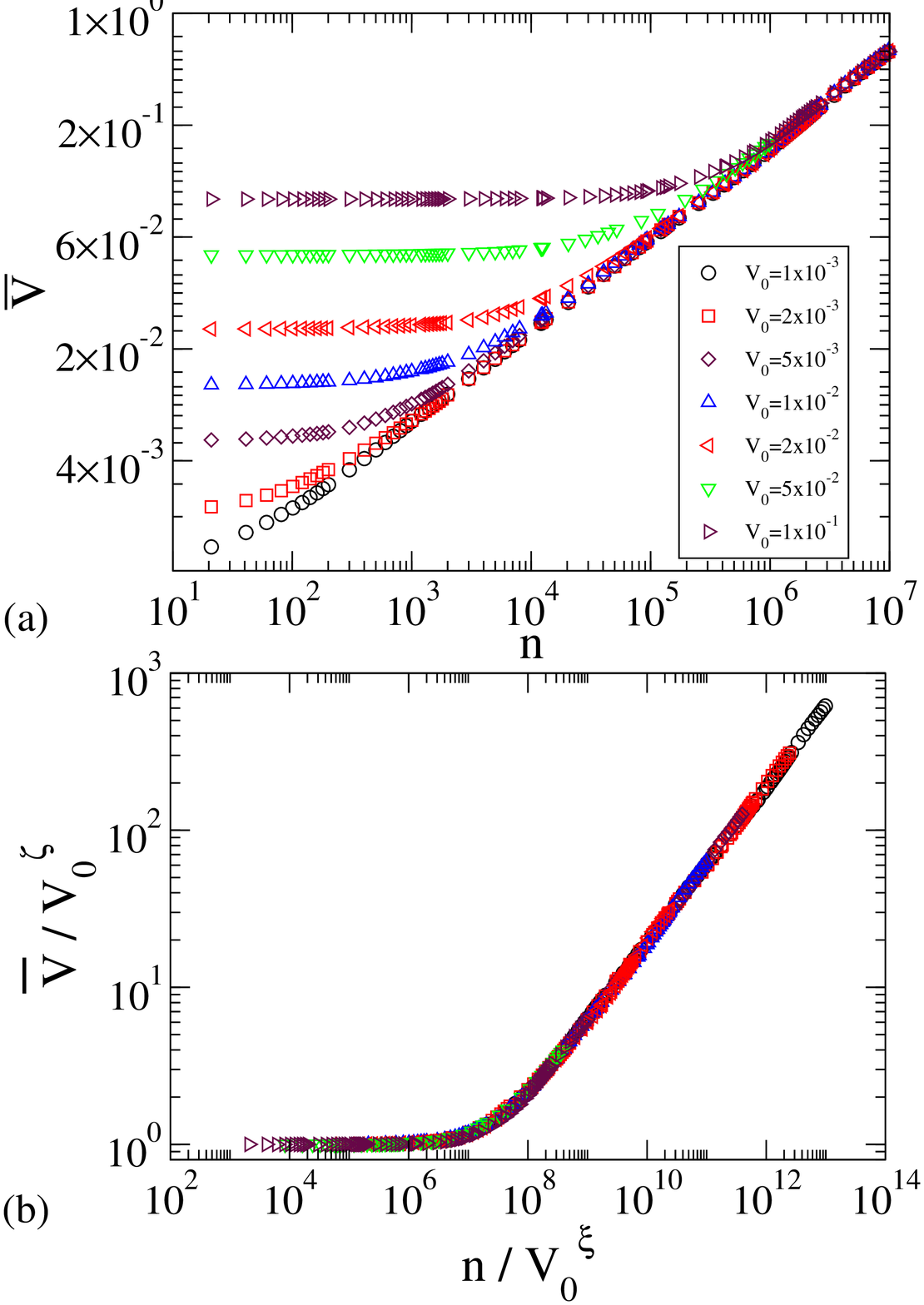}}
\caption{\it{(a) Behaviour of average velocity for different values
of $V_0$; (b) their collapse onto a single and universal plot.}}
\label{fig21}
\end{figure}

Note that the scaling exponents are determined if the critical
exponents $\zeta$ and $\nu$ were numerically obtained. The exponent
$\nu$ is obtained from a power law fitting for the average velocity
when $n\gg{n_x}$. Thus, an average of these values gives
$\nu=0.49(1)$.  Figure \ref{fig20} shows the behaviour of (a),
${\bar V}_{sat} \times V_0$ and (b), $n_x\times V_0$. Applying power
law fittings we obtain $\zeta=1.00(1)\cong 1$ and $\xi=2.01(3)$.
Considering the previous values of both $\zeta$ and $\nu$  and using
$\xi={\zeta/\nu}$, we find that $\xi=2.04(2)$. Such result indeed
agrees with our numerical data. In order to confirm the initial
hypotheses and, since the values of the scaling exponents $\zeta$,
$\nu$ and $\xi$ are now known, we will collapse all the curves onto
a single and universal plot, as demonstrated in Fig. \ref{fig21}.
Additionally, such result allow us to confirm the validity of LRA
conjecture since the $\bar {V}$ grows unbounded for $n>>n_x$.

\subsection{Scaling results for the dissipative case.}

\begin{figure}[t]
\centerline{\includegraphics[width=0.60\linewidth]{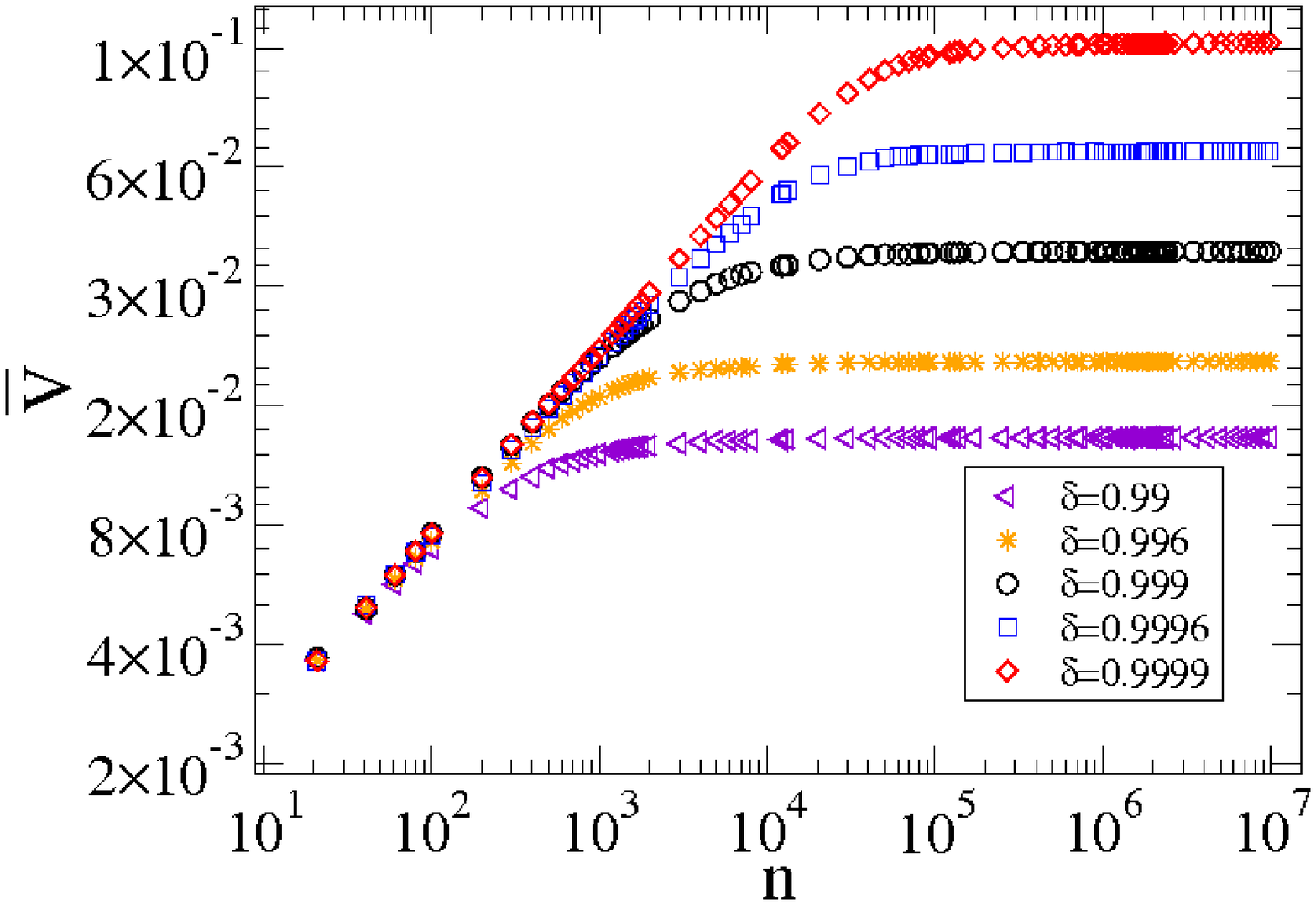}}
\caption{\it{Behaviour of $\overline{V} \times n$ for different
values of $\delta$, as labeled in the figure.}}
\label{fig8}
\end{figure}

\begin{figure}[t]
\centerline{\includegraphics[width=0.60\linewidth]{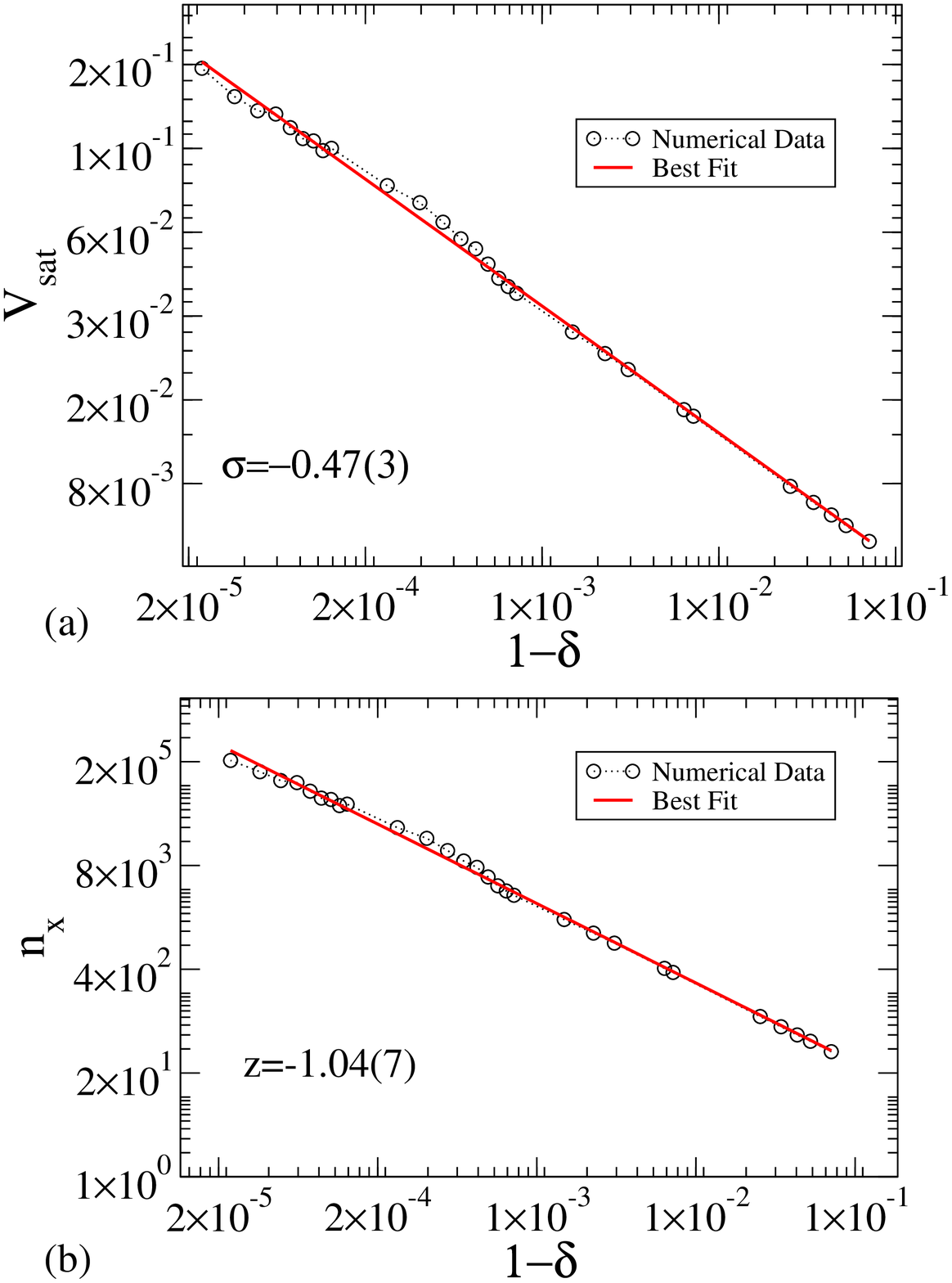}}
\caption{\it{(a) Behaviour of $\overline{V}_{sat} \times {(1-\delta)}$.
(b) Behaviour of the crossover number $n_x$ against $(1-\delta)$. A
power law fitting in (a) furnishes $\sigma=-0.47(3)$ while in (b)
$z=-1.04(7)$.}}
\label{fig9}
\end{figure}

In this section we will use the same scaling formalism as used in the previous section. 
We concentrate to characterize the behaviour of the
average velocity in terms of the number of collisions with the boundary
and as a function of the damping coefficient along the normal component of the
particle's velocity, $\delta$. 
We study a dissipative version of the Lorentz Gas close to
the transition from unlimited to limited energy growth. Indeed, such a
transition happens when the control parameter $\delta\rightarrow 1$ and 
it is better characterized if adopt the following transformation $\delta\rightarrow (1-\delta)$. To
obtain the average velocity, each initial condition has a fixed initial
velocity, $V_0=10^{-4}$ and randomly chose $t \in [0, 2\pi]$, $\theta \in [0, 2\pi]$ and $b \in [1-(\epsilon_x+\epsilon_y), -1+(\epsilon_x+\epsilon_y)]$. The control parameter $\gamma$
were fixed as been $\gamma=1$.

It is shown in Fig. \ref{fig8} the behaviour of average velocity as function of the number of collision for different values of $\delta$. Note that, 
for different values of $\delta$, the average velocity, for small $n$, starts to grow with the same slope and them they bend towards a regime of saturation
for long enough values of $n$. The changeover from growth to
the saturation is marked by a typical crossover number $n_x$. For
such a behaviour, we can propose the following scaling hypotheses:
\begin{enumerate}
 \item {When $n\ll{n_x}$ the average velocity is 
\begin{equation}
\overline{V}\propto n^{\eta},
\label{eq17}
\end{equation}
}
\item{For long time, $n\gg{n_x}$, the average
velocity approaches a regime of saturation, that is described as
\begin{equation}
\overline{V}_{sat}\propto (1-\delta)^{\sigma},
\label{eq018}
\end{equation}}
\item{The crossover number that marks the regime of
growth to the constant velocity is written as
\begin{equation}
n_x\propto (1-\delta)^{z}~,
\label{eq019}
\end{equation}
where $\sigma$, $\eta$ and $z$ are critical exponents.}
\end{enumerate}

\begin{figure}[t]
\centerline{\includegraphics[width=0.60\linewidth]{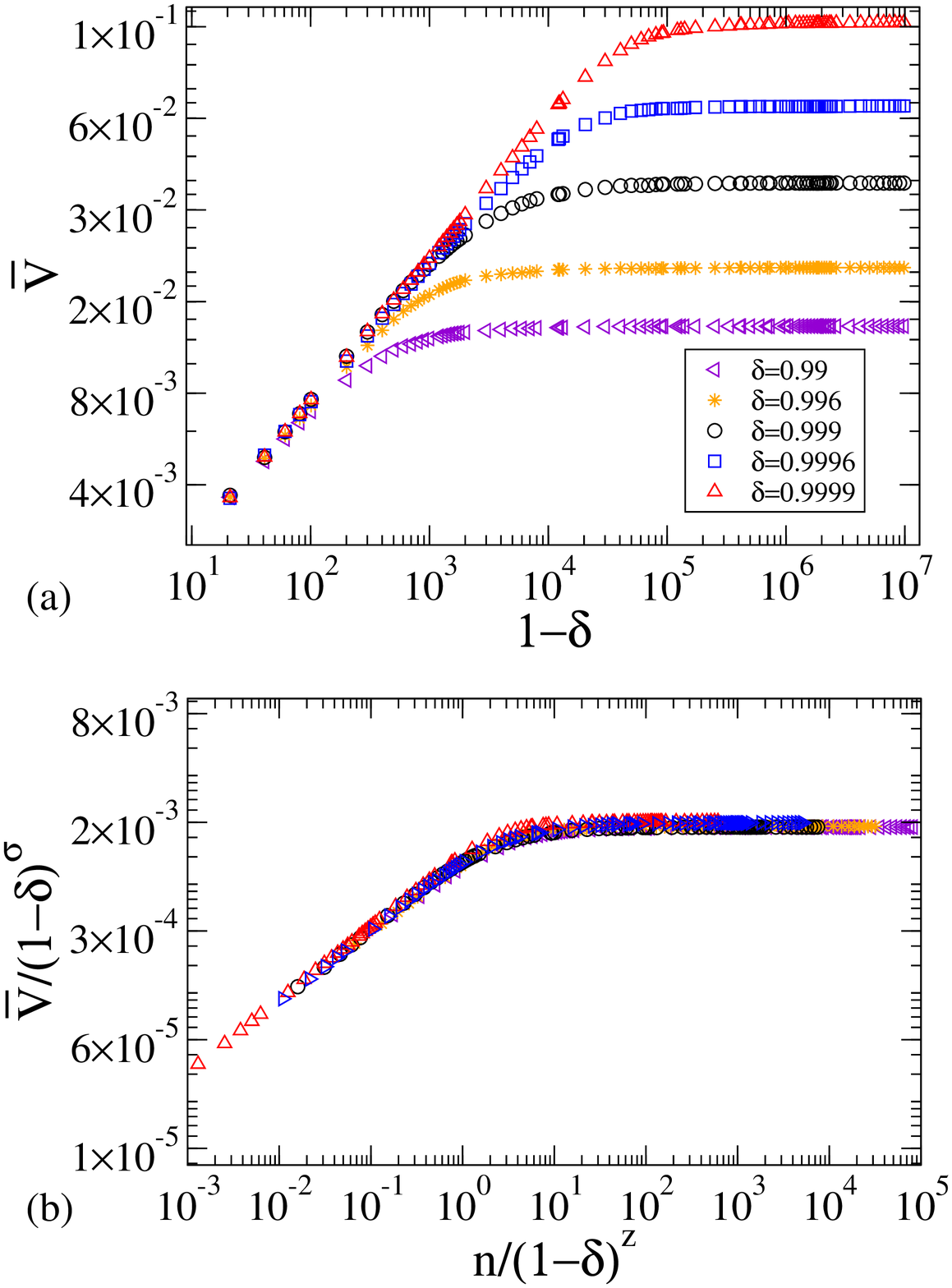}}
\caption{(a) Different curves of the $\overline{V}$ for five different control parameters. (b) Their collapse onto a single and universal plot.}
\label{fig10}
\end{figure}

These scaling hypotheses allow us to describe the average velocity in terms of a scaling function of the type

\begin{eqnarray}
\overline{V}[n,(1-\delta)]=l\overline{V}[{\rm l}^p{n},{\rm l}^q{(1-\delta)}]~,
\label{eq720}
\end{eqnarray}
where {\it p} e {\it q} are scaling exponents and ${\rm l}$ is a scaling factor. Since ${\rm l}$ is a scaling factor, we can chose it such that ${\rm l}^{p}n=1$, yielding
\begin{eqnarray}
\overline{V}[n,(1-\delta)]={n}^{-1/p}\overline{V}_1[(n)^{-q/p}(1-\delta)]~,
\label{eq721}
\end{eqnarray}
where $\overline{V}_1[(n)^{-q/p}(1-\delta)]=\overline{V}[1,(n)^{-q/p}(1-\delta)]$ is assumed to be constant for $n\ll{n_x}$. 
Comparing Eqs. (\ref{eq17}) and (\ref{eq721}), we obtain $\eta=-1/p$. A power law fitting gives us that $\eta=0.471(2)$. Such value
was obtained from the range of $\delta\in[0.99,0.99999]$. Choosing now ${\rm l}^q(1-\delta)=1$, we have that ${\rm l}=(1-\delta)^{-1/q}$ 
and Eq. (\ref{eq720}) is rewritten as
\begin{eqnarray}
\overline{V}[n,(1-\delta)]=(1-\delta)^{-1/q}\overline{V}_2[(1-\delta)^{-p/q}n]~,
\label{eq722}
\end{eqnarray}
where $\overline{V}_2[(1-\delta)^{-p/q}n]=\overline{V}[(1-\delta)^{-p/q}n,1]$ is assumed to be constant for $n\gg{n_x}$. 
Comparing Eqs. (\ref{eq018}) and (\ref{eq722}), we obtain $-1/q=\sigma=-0.47(3)$ [see Fig. \ref{fig9} (a)]. 
Using now the expressions obtained for the scaling factor $l$,  we can easily show that
\begin{eqnarray}
z={\sigma\over\eta}=0.99(5)~,
\label{eq23}
\end{eqnarray}
which is quite close to the value obtained numerically, as shown in Fig.
\ref{fig9} (b). A confirmation of the initial hypotheses is made by a
collapse all the curves of ${\bar V}\times n$ onto a single and
universal plot, as shown in Fig. \ref{fig10}, showing that the system is scaling invariant. We also shown that
dissipation causes a drastic change in the behavior of $\overline{V}$. Note that when $ \delta \rightarrow 1$,
implies that Eq. \ref{eq018}  and Eq. \ref{eq019} diverge, thus recovering the results for the
conservative case, i.e., Fermi acceleration. However, when $\delta$ is
slightly less than 1, the average velocity grows and then reach a regime of saturation for long enough time. 
Our results reinforce that dissipation introduced via damping coefficients is a sufficient condition to suppress the phenomenon of 
Fermi acceleration.

\section{Conclusion}
\label{sec4}

In this paper we consider the problem of a classical Lorentz Gas
considering both the static and time-dependent boundary. For the
static case we obtain the mapping that describes the dynamics of the
system and we have shown that the model has a chaotic component
characterized with positive Lyapunov Exponent. After that, we
introduce a new type of the time-dependent perturbation on the
boundary. Our results confirm the validity of LRA conjecture for the conservative case since
the phenomenon of Fermi acceleration is observed. When dissipation, via damping coefficients, 
is introduced into the model we observed that the average velocity
grows with time and then reaches a constant value for large enough time 
confirming that Fermi acceleration is suppressed.  Finally, the
average quantities were described by scaling functions with
characteristic exponents whose validity was confirmed with the
collapse of the curves of $\bar {V}$ into an universal plot.

\section*{ACKNOWLEDGMENTS}
D.F.M.O gratefully acknowledges Max Planck Institute for financial support. E. D. L. is grateful
to FAPESP, CNPq and FUNDUNESP, Brazilian agencies. The authors acknowledge Dr. Douglas Fregolente
for a careful reading on the manuscript.

\end{document}